\documentclass{article}

\usepackage[title]{appendix}
\usepackage{arxiv}
\usepackage{color}
\usepackage[utf8]{inputenc} 
\usepackage[T1]{fontenc}    
\usepackage{hyperref}       
\usepackage{url}            
\usepackage{booktabs}       
\usepackage{amsfonts}       
\usepackage{nicefrac}       
\usepackage{microtype}      
\usepackage{lipsum}
\usepackage{graphicx}
\usepackage{caption}
\usepackage{subcaption}
\graphicspath{ {./images/} }
\usepackage{biblatex}
\title{Physics-Constrained Generative Adversarial Networks for 3D Turbulence}

\author{
  Dima Tretiak\\
  CCS-2/Computational Physics and Methods\\
  Los Alamos National Laboratory\\
  Georgia Institute of Technology\\
  \texttt{dtretiak@gatech.edu} \\
  
   \And
 Arvind T. Mohan \\
  CCS-2/Computational Physics and Methods\\
  Center for Nonlinear Studies\\
  Los Alamos National Laboratory\\
  \texttt{arvindm@lanl.gov} \\
  
  \And
 Daniel Livescu \\
  CCS-2/Computational Physics and Methods\\
  Los Alamos National Laboratory\\
}

\begin{document}
\maketitle

\begin{abstract}
Generative Adversarial Networks (GANs) have received wide acclaim among the machine learning (ML) community for their ability to generate realistic 2D images. ML is being applied more often to complex problems beyond those of computer vision. However, current frameworks often serve as “black boxes” and lack physics embeddings, leading to poor ability in enforcing constraints and unreliable models. In this work, we develop physics embeddings that can be stringently imposed, referred to as hard constraints, in the neural network architecture. We demonstrate their capability for 3D turbulence by embedding them in GANs, particularly to enforce the mass conservation constraint in incompressible fluid turbulence. In doing so, we also explore and contrast the effects of other methods of imposing physics constraints within the GANs framework, especially penalty-based physics constraints popular in literature. By using physics-informed diagnostics and statistics, we evaluate the strengths and weaknesses of our approach and demonstrate its feasibility.
\end{abstract}

\section{Introduction}

Turbulence is a chaotic phenomenon that can be described in many applications by a system of nonlinear Partial Differential Equations (PDEs), but for which no general analytic solution has been found. The few analytical solutions that exist are narrow in scope and only applicable to specific problems. Numerical solutions also exist, but are limited by computational power and do not apply to all cases. Furthermore, the multi-scale nature of turbulence greatly increases the complexity of the problem and limits the scales at which numerical solutions can be resolved, given current computational constraints. Most practical flows, however, are turbulent and range from climate to aerodynamics to astrophysics and beyond.  Therefore, efficient and accurate modeling of turbulence has the potential to impact a wide variety of applications. Because of the highly nonlinear nature of turbulent flows, machine learning (ML) methods, especially deep learning (DL), have seen a large interest lately. DL has become a popular strategy to tackle many complex fluids problems \cite{Moghaddam, Zhang}. Particularly, deep learning architectures such as Generative Adversarial Networks (GANs)\cite{Goodfellow} and Convolutional neural networks~\cite{lecun1995convolutional} have been attractive for modeling turbulence since they can handle high-dimensional spatio-temporal datasets. An attractive architecture that uses stacked convolutional layers in GANs is the Deep Convolutional GAN, or DCGAN~\cite{radford2015unsupervised}. DCGANs can model 3D turbulence~\cite{Xie,Mohan_JoT,mohan2019compressed}, but they fall into the category of “black-box” models since they are unaware of the relevant physics. These physics-agnostic models fare poorly when required to capture key constraints such as conservation laws~\cite{Mohan_constraints}. Furthermore, their results are exceedingly difficult to interpret as neural networks get deeper and reach increasing levels of abstractions. Although a neural network can learn an optimized model that produces outputs with low error during training, they do not guarantee the solutions are physically consistent. 

Therefore, considerable focus has been on introducing physics into ML models by constraining the neural networks with domain-specific knowledge, so that the model converges to a space where its predictions not only have low error but also satisfy key physics. Two broad strategies have arisen in literature: soft and hard constraints implementations. The more common of the two, the soft constraint approach, typically consists of adding terms to the loss function(s) of a model to enforce a specific constraint. Hard constraint approaches, on the other hand, are embedded directly into the architecture of the network and automatically produce predictions satisfying physics constraints. Therefore, hard constraints are explicitly satisfied during training. However, neural networks with hard constraints introduce significant complexity, since the core architecture needs redesign to be simultaneously consistent with both backpropagation and physical constraints. Some notable approaches for hard constraints implementations include those from Jiang \cite{Jiang} which involve enforcing constraints through differentiable PDE layers, and the embedded invariance approach by Ling et al.~\cite{Ling}. There are several works that use soft constraints implementations, such as Wu et. al. \cite{Wu} who employ a generative model and enforce statistical constraints with regularization of the loss function. Wang et al. \cite{Wang} take a hybrid approach and introduce a Machine Learning Reynolds-Averaged Navier-Stokes (ML-RANS) coupled model which takes advantage of spectral filters to perform physics-informed DL, while also embedding physical constraints in the loss function.

A widely used soft-constraints approach is the Physics-Informed Neural Networks (PINNs) strategy by Raissi et al. \cite{Raissi} which was introduced for nonlinear PDE-based systems, and several variants of PINNs have been proposed for various applications~\cite{pang2019fpinns,cai2022physics,kharazmi2019variational}.

The focus of this work is on developing hard constraint methods for the GANs architecture as applied to 3D homogeneous isotropic turbulence, expanding on work from Mohan et al.\cite{Mohan_JoT}. We introduce two different approaches to enforcing hard constraints in DCGANs that enforce the divergence-free nature of incompressible flow to machine precision for the specific method considered. This physics-embedded DCGAN, or \textit{PhyGAN} network not only models the detailed statistical properties of 3D turbulence, but also does so while strictly adhering to the mass conservation law. We employ rigorous physics-based statistical diagnostics to analyze the strengths and weaknesses of our approach. Finally, we also compare the hard constraint results with those obtained using a soft-constraints implementation, and comment on the key differences.

\section{Training Data} \label{Data}
We trained our networks using 96 snapshots of Direct Numerical Simulation (DNS) of stationary homogeneous isotropic turbulence (HIT) data, produced through a standard fully dealiased pseudo-spectral method ~\cite{Livescu00,livescu2011direct,Daniel} by solving the incompressible Navier-Stokes equations together with the incompressibility condition shown in Equation (\ref{NS}) on a $128^3$ grid. Equation (\ref{NS}) is shown in index notation, where $v_i$ is the velocity component in direction $i$, $p$ is the pressure, $\nu$ the kinematic viscosity, and $f_i^v$ is linear forcing term that ensures statistical stationarity~\cite{Lundgren,Daniel}. The forcing acts only at small wavenumbers, $k<1.5$, to maximize the Reynolds number for the given mesh. The Reynolds number based on the Taylor microscale is $Re_\lambda \sim 91$.

\begin{equation}
\partial_{t}v_{i}+v_{j}\partial_{x_{j}}v_{i}=-\frac{{1}}{\rho}\partial_{x_{i}}p+\nu\Delta v_{i}+f_{i}^{v},\qquad \partial_{x_{i}}v_{i}=0,
\label{NS}
\end{equation}

\begin{figure}[h]
	\centering
	\includegraphics[width=150mm]{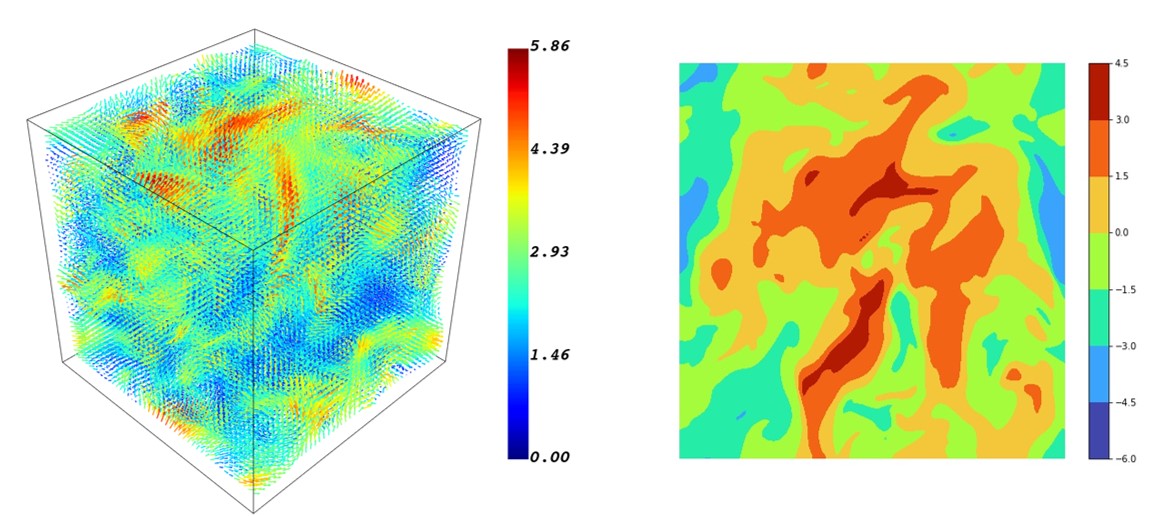}
	\caption{Visualization of the 3D ground truth velocity flow field $\mathbf{v}$ with vector direction shown as quivers and magnitude (total kinetic energy) indicated by color (left). Every third vector in each axis is plotted for clarity. The panel on the right shows a contour plot of a 2D slice of the magnitude of the $x$ component of velocity $\mathbf{v}$ in the $y$-$z$ plane. 
	} 
	\label{DNS Figures}
\end{figure}

Each snapshot used for learning is a 3D volume of size $128^3$ and contains the three components of velocity. The snapshots capture $\approx 3$ eddy turnover times after the flow reaches stationarity. The statistically stationary data are ideal for DCGANs that predict individual snapshots, since temporal correlations are not modeled. Therefore, our primary goals for our DCGAN architecture are to model the relevant statistics of the flow and reinforce the divergence-free hard constraint $\nabla \cdot \mathbf{v} = 0$. Figure \ref{DNS Figures} depicts the 3D flow field and a 2D slice of the field, for illustration. 

\subsection{Diagnostics}\label{Diag}
The compare the accuracy of the NN models, we use three physics-based diagnostics \cite{King}. The first diagnostic is a comparison of the energy spectrum with the ground truth. The energy spectrum, $E(k)$, describes the relationship between the inverse of the wavelength (denoted as Fourier wavenumber $k$) and the kinetic energy contained by the respective Fourier mode. Hence, the energy spectrum is a useful tool in evaluating a model's ability to capture the energy content at various length scales.

The next diagnostic is the probability distribution function (PDF) of the longitudinal velocity gradient $Z$. Small scale turbulence is highly intermittent, as indicated by the non-Gaussian tails and non-zero skewness \cite{pope2000turbulent} of the $Z$ PDF. The non-zero skewness is a consequence of the three-dimensional structure of turbulence, which is presented as the next diagnostic. 

Lastly, the most complex diagnostic tool considered here is the joint PDF of the second ($Q$) and third ($R$) invariants of the anisotropic part of the filtered velocity gradient tensor at three different length scales: small ($r=1$), inertial ($r=8$), and large ($r=32$), where $r$ roughly corresponds to the grid spacing~\cite{chertkov1999lagrangian}. The Q-R plane at each $r$ value allows for the evaluation of the structure of the flow at that scale, since each quadrant corresponds to a specific topology. In the inertial range, 3D turbulent flows exhibits a ``teardrop" shape in the Q-R plane. This shape is a consequence of the vortex stretching mechanism for generating vorticity. 

\section{Deep Convolutional Generative Adversarial Networks for 3D Turbulence}

\label{DCGAN Section}

\begin{figure}[h]
	\centering
	\includegraphics[width=120mm]{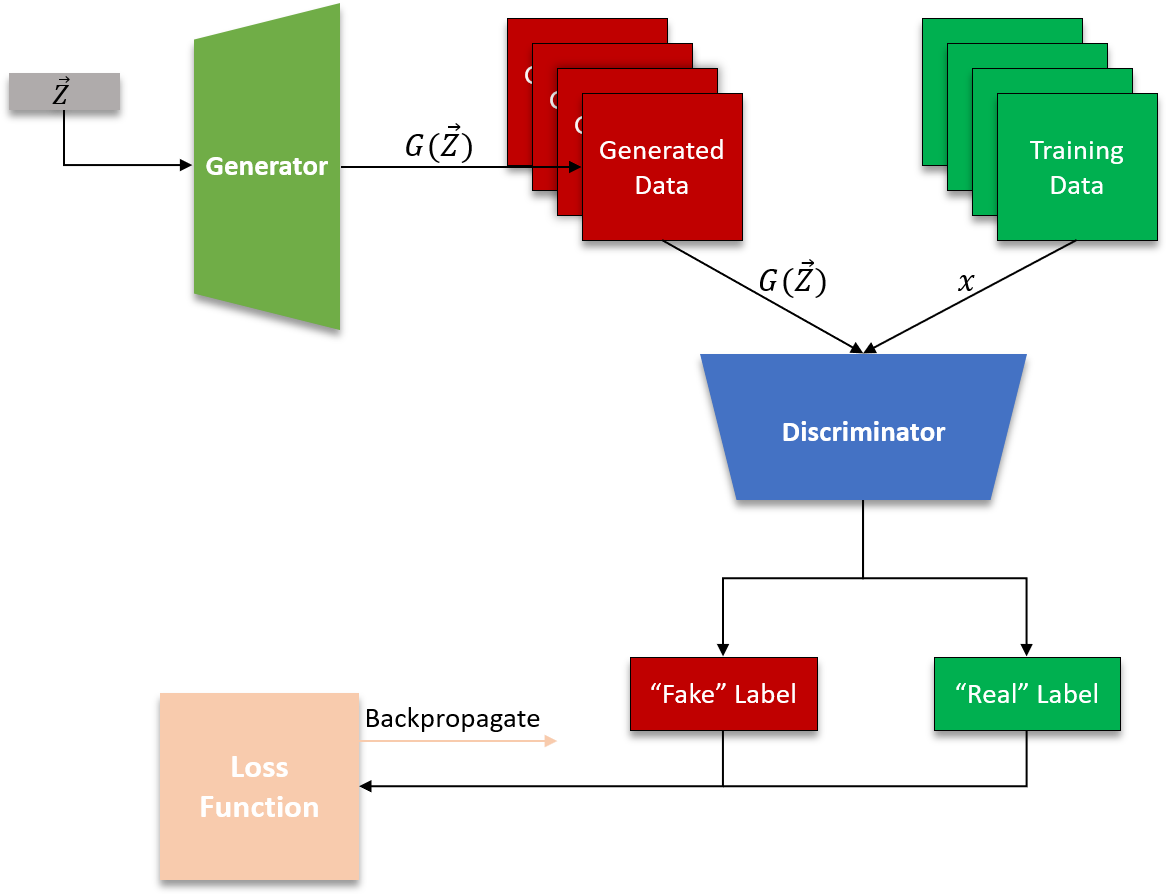}
	\caption{General architecture of a DCGAN with data flowing between the Generator and Discriminator networks.}
	\label{DCGAN Architecture}
\end{figure}

Deep Convolutional Generative Adversarial Networks, or DCGANs, comprise two convolutional neural networks that compete against each other in a min-max game. Figure \ref{DCGAN Architecture} illustrates the general architecture of a DCGAN model. The Generator network, in green, upsamples a random latent vector $z$ into a sample of data, $G(z)$, of equivalent size to the training data, $x$. The Discriminator network, in blue, has a domain equivalent to the range of the Generator and serves as a binary classifier. It labels the data as “real” or “fake”; the “real” label corresponds to the ground truth training data, whereas the “fake” label corresponds to the data from the Generator network. Hence, the Generator network’s objective is to minimize the difference between the Discriminator’s predication on the generated data, $D(G(z))$, and the Discriminator’s predication on the training data $D(x)$. In other words, the Generator seeks to “fool” the Discriminator into misclassifying the generated data as “real.” By doing so, the Generator models the distribution of the ground truth training data. 

In GANs, both networks are trained concurrently in the same training loop. Each iteration updates the Discriminator followed by an update to the Generator. While updating one network, the other network’s parameters are frozen (i.e. if the Generator’s weights are being updated, the Discriminator’s weights are frozen). Hence, there is a balance between the networks such that if one becomes too accurate relative to the other, meaningful gradients can no longer be propagated through the networks.

The DCGAN framework can be described by the objective function $V(D,G)$, given by Equation (\ref{GANS}). The Generator’s objective is to minimize the function $V$, while the Discriminator’s objective, on the other hand, is to maximize the function $V$. The Generator is a mapping from the latent variable $z$, drawn from a prior distribution, $p_z$, to the data space. The Discriminator is a scalar function of data space that outputs probability that input
was genuine (i.e. drawn from true data distribution, $p_{data}$). Here,  the “real” label is represented with a $1$ and the “fake” label with a $0$. $V(D,G)$ is given by the sum of expected value of the log of the Discriminator predicting that Generator's generated data is fake and the expected value of the log of the Discriminator predicting that real-world data, $x$, is real. The Generator cannot directly affect the $\log D(\mathbf{x})$ term in the function, so, for the Generator, minimizing the loss is equivalent to minimizing $\log (1 - D(G(\mathbf{z}))$.
This objective function sets up the adversarial game between the two networks and is satisfied only when the Generator has reproduced the training data distribution. As the Generator continues to improve, the two expectation terms in Equation (\ref{GANS}) become $\log(0.5)$ as the Discriminator can no longer distinguish between real and generated samples. In practice, reaching the networks’ objectives is often difficult because of vanishing gradients; hence, DCGANs have become notorious for unstable and unproductive training \cite{Wiatrak} and often need supplemental improvements to improve training stability. A common failure mode occurs when the Discriminator becomes too accurate, resulting in vanishing gradients for the Generator as the Discriminator labels become close to exactly 1 or 0. More information is given in Appendix \ref{label_smoothing}.

\begin{equation}
V(D,G) = \mathbb{E}_{\mathbf{x}\sim p_{data}(\mathbf{x})}[\log D(\mathbf{x})] + \mathbb{E}_{\mathbf{z}\sim p_z(\mathbf{z})}[\log (1 - D(G(\mathbf{z}))]
    \label{GANS}
\end{equation}

Notably, the original DCGAN architecture was designed for use in 2D image generation. However, for 3D turbulence, Mohan et. al. \cite{Mohan_JoT} utilized several adaptations proposed for DCGANs such as 3D convolutions, modifications for stability improvements, resize convolutions, and variable kernel sizes in order to capture statistics of 3D flow (see Appendix \ref{GANs Details}). 

\begin{figure}[h]
	\centering
	\includegraphics[width=100mm]{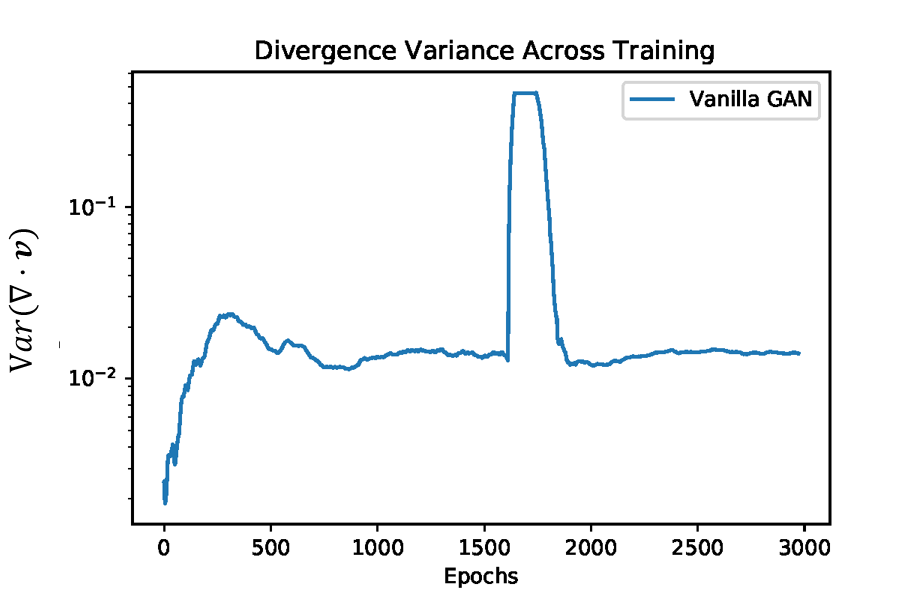}
	\caption{Plot of the velocity divergence variance (computed via second order central finite differences) across training epochs. The divergence variance $\langle\nabla \cdot \mathbf{v} - \langle\nabla \cdot \mathbf{v}\rangle \rangle^2$, where square brackets denote volume averages, rather than the mean $\langle\nabla \cdot \mathbf{v}\rangle$, is plotted on the Y-axis in order to evaluate the local departures from the constraint. 
    }
	\label{DivVanilla}
\end{figure}

We utilize the aforementioned model as a baseline "Vanilla" DCGAN (i.e. the model has no physics) for comparison with PhyGAN. Although the diagnostics illustrated by Mohan \cite{Mohan_JoT}
show that the Vanilla DCGAN model performed satisfactorily without any enforcement of constraints, it fares poorly in satisfying the ``divergence-free" incompressibility condition  $\nabla \cdot \mathbf{v} = 0$. We illustrate this in Figure \ref{DivVanilla} where the variance of the velocity divergence is plotted as a function of training epochs. We see that the velocity divergence is not only fluctuating during the course of training, but also gradually increases from the known ground truth value of 0, despite the model predictions capturing the diagnostic chosen. This highlights the poor interpretability of such black-box models since we can get reasonably accurate models for the wrong reasons. Therefore, our development of hard constraints not only improves the model's accuracy but also enhances our ability to trust the model's behavior based on physical metrics, with less dependence on hyperparameter choices.

\section{Physics Embedding in DCGANs}
 
\subsection{Soft Constraints} \label{SC section}

As a basis for comparison with hard constraints implementations, we first implement soft constraints through modification of the loss function. Given that the generator's goal is to minimize the value function shown in Equation \ref{GANS}, a divergence term $\nabla \cdot \mathbf{v}$ can be added such that the flow divergence is minimized in conjunction with the loss. Furthermore, the divergence term is regularized by a parameter $\lambda$ that serves to assign relative importance to the incompressiblity condition; we found $\lambda = 1$ to be suitable in our tests, although this may change for other datasets/models. The value function was also adjusted to reflect a "real" label of 1 and a "fake" label of zero, since we found swapping these labels achieved greater stability in training. Thus, the generator's new objective is to minimize the value function shown in Equation \ref{SCLoss}. In practice, we used a binary cross entropy loss function with swapped labels to implement this objective.  

\begin{equation}
    V(G) = \mathbb{E}_{\mathbf{x}\sim p_{data}(\mathbf{x})}[\log(1- D(\mathbf{x}))] + \mathbb{E}_{\mathbf{z}\sim p_z(\mathbf{z})}[\log D(G(\mathbf{z}) + \lambda(\nabla \cdot \mathbf{v})]
    \label{SCLoss}
\end{equation}

In order to calculate the velocity divergence during training, we implemented a finite difference scheme in PyTorch~\cite{Paszke}. The scheme consists of 2nd order central differences along the interior points and one-sided second order forward/backward approximations at the boundaries \cite{Lele92}. The library's autograd system allows us to backpropogate through the divergence operation. Hence, we are able to minimize the divergence in conjunction with the loss.
 
\subsection{Hard Constraints}
We now present the design of our hard physics constraints with two different approaches. While the physics embeddings can be injected in any network with a convolutional architecture, in GANs we use the Generator since it makes the final predictions. Thus, we take inspiration from Mohan et. al. \cite{Mohan_constraints} and implement the physics embeddings into the Generator. The key idea is to exploit the Helmholtz decomposition $V \,=\, \nabla \times A + \nabla \psi$, where $V$ is the velocity in an incompressible fluid and $A$ is the vector potential. If the velocity satisfies this relationship, then we automatically satisfy $\nabla \cdot V = 0$. In this work, we utilize the homogeneous isotropic turbulence flow dataset, where the boundary conditions are periodic and therefore $\psi = 0$, simplifying the final relationship to $V \,=\, \nabla \times A$. In essence, we transform the problem of learning $V$ to first learning an appropriate vector potential $A$ as an intermediate solution, followed by taking the curl to obtain the predicted velocity $\tilde{V}$. For hard constraint implementations, the loss function is a standard MSE loss function between predicted velocity $\tilde{V}$ and true velocity from the training data $V$. 

\begin{figure}[h]
	\centering
	\includegraphics[width=165mm]{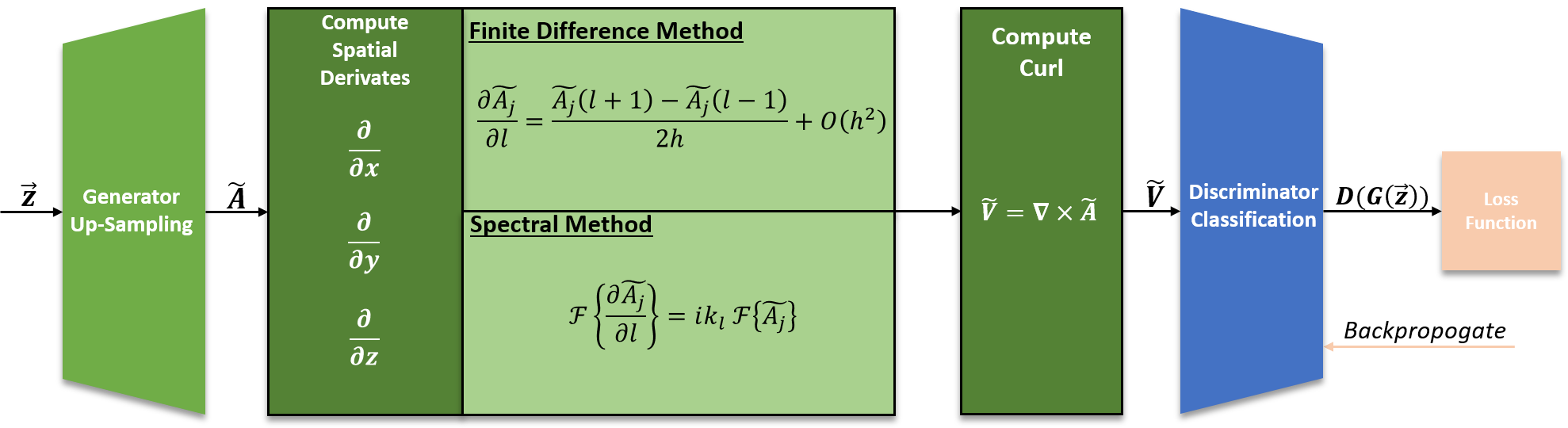}
	\caption{Hard constraints schematic depicting Generator creating the backpropagation-compatible physics layers designed to enforce the incompressiblity condition. First, the random vector $z$ is passed to the Generator which produces a potential vector $\widetilde{A}$. The potential vector is then passed through the non-trainable physics layers where the curl is computed to create the final vector field $\widetilde{V}$. Second order central finite differencing for the potential vector component $A_j$ in direction $l$ with grid spacing $h$ is shown in the top box and spectral differencing in the bottom box, where $k_l$ is the wavenumber vector in direction $l$ and $i=\sqrt{-1}$. 
    }
	\label{HCSchematic}
\end{figure}

Figure \ref{HCSchematic} illustrates the general architecture of how the current hard constraint is implemented, with both approaches shown in the same schematic for brevity. First, the upsampling layers in the DCGAN predict an intermediate solution, which is then fed into layers that compute partial derivatives. These partial derivatives are calculated so that the next layer can compute the curl on the output of the upsampling layers, to output the velocity prediction $\tilde{V}$. Finally, the discriminator checks the solution $\tilde{V}$ and proceeds to the loss function. The key aspect of this architecture is that all layers are compatible with backpropagation and do not impede the training of the learnable parameters. This ensures that the DCGAN training is intimately aware of the constraints at an architectural level in both forward and backward passes, and not just during the loss computation, unlike soft constraints. The difference between the two approaches lies in how the derivatives are computed, which we now discuss. 

In contrast to \cite{Mohan_constraints}, we forego the usage of a finite difference kernel, and directly apply the same finite difference operation mentioned in section \ref{SC section}. Enforcing the incompressiblity condition shown in Equation (\ref{NS}) in GANs is the primary focus of this work. Given that we work with discrete data, we employ two primary methods to approximate the necessary partial derivatives: finite differences and spectral differentiation. The former method is used in the soft constraints variant and one of the hard constraints variants, while the latter in the other hard constraints variant. The end results of these three implementations are then compared side-by-side.

\subsection{Finite Difference} \label{finite_difference}

In constructing a backpropagation compatible derivative operator, we first build off the Numpy implementation of gradient \cite{numpy} in order to calculate the finite difference approximations. We utilize PyTorch Tensors in conducting the calculations in order to take advantage of the package's Autograd capability. Similar to the soft constraints implementation, we use second order central finite differences at the interior points and second order one-sided forward/backward approximations at the boundary points.

\subsubsection{Hard Constraints: Spectral Generator Embedding}
The spectral Generator embedding takes a similar approach to the finite difference embedding: compute $\mathbf{\widetilde{A}}$, take the curl, and automatically generate a divergence-free field. Unlike the previous method, this method utilizes the Fourier transform of the velocity field and uses spectral derivatives, which are just multiplications in Fourier space as shown Figure \ref{HCSchematic}, where $k_l$ is the wavenumber in direction $l$ and $\mathcal{F}$ denotes the Fourier transform. This transform is widely used in the study of turbulence, as the spectral information is related to the different scales represented in the flow; smaller wavenumbers correspond to large scale structures, whereas larger wavenumbers correspond to small scale structures. 

Due to the data generation method, spectral differencing is directly consistent with how the derivatives are computed in the ground truth dataset, as opposed to the finite difference implementation. We also note related work which used trainable Fourier embeddings in a neural network by Li et al.\cite{Li}, although their focus was operator learning on canonical PDEs and laminar flow problems.

\section{Experiments}

In this section, we perform numerical experiments in training with both hard and soft constraints, including two different types of hard embeddings, on the homogeneous isotropic turbulence dataset described in Section~\ref{Data}. The details of the architecture implementation and training hyper-parameters can be found in Appendix~\ref{GANs Details}. In all the experiments, a key quantity of interest is the change in velocity divergence of the predicted flow across training epochs. This is different from the conventional loss vs epochs plots, which only indicates prediction accuracy and not fidelity to constraints. As hypothesized in Mohan~\cite{Mohan_constraints}, a correctly imposed hard constraint differs from soft constraint implementation in fidelity to the respective constraint at not just during the final epoch, but also intermediate epochs. In other words, hard constraints implementations must show stronger adherence to the given constraints during training since the architecture only permits such solutions, and should outperform soft constraints implementations. We will analyze the experiments from this standpoint, and also assess the accuracy of the results using the physics-based diagnostic tests presented in Section~\ref{Diag}.

\subsection{Soft Constraints}
Figure \ref{DivSC} below illustrates the progression of the velocity divergence variance across epochs during training. When compared to the Vanilla GAN model, the soft constraints model shows clear, albeit noisy improvement. The quasi-periodic noise is likely a result of the use of cyclic learning rates as described in Appendix \ref{lr}. However, due to the presence of the physical constraint in the neural networks, this value decreases overall during training. In contrast, velocity divergence variance rises in the Vanilla GAN in the absence of any physical information.

\begin{figure}[h]
	\centering
	\includegraphics[width=100mm]{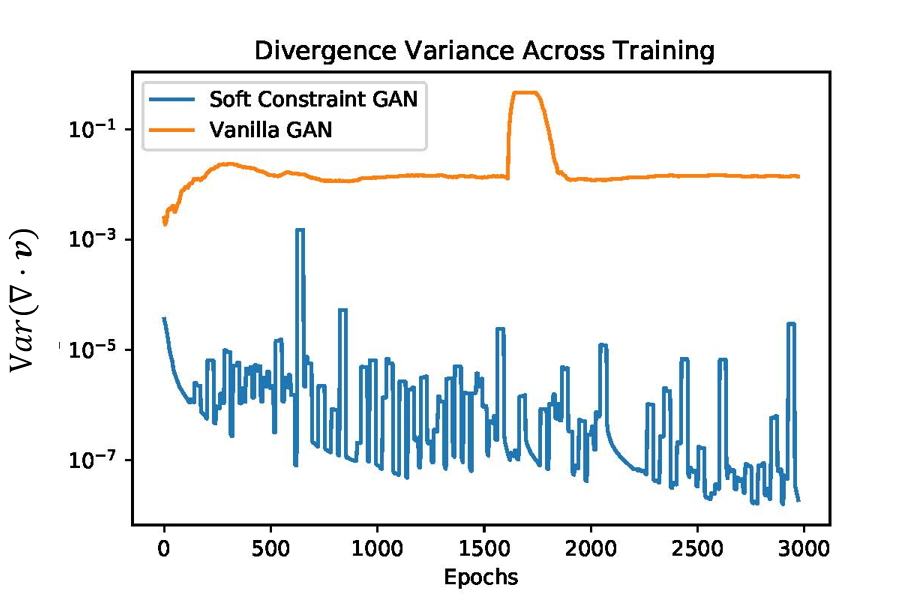}
	\caption{Velocity divergence variance plotted as the soft constrained model trains. Given that the divergence loss parameter is calculated through finite differencing, so is the velocity divergence in the plot.}
	\label{DivSC}
\end{figure}

Notably, this method is relatively simple and can be easily implemented for other constraints and other network architectures. Essentially, the soft constraints technique can be applied to any predictive model that involves a loss function minimization. This is particularly convenient for a wide variety of physical problems involving differential constraints in PDEs and ODEs. 

\begin{figure}
	\centering
	\begin{subfigure}[b]{0.49\textwidth}
	    \centering
		\includegraphics[width = \textwidth]{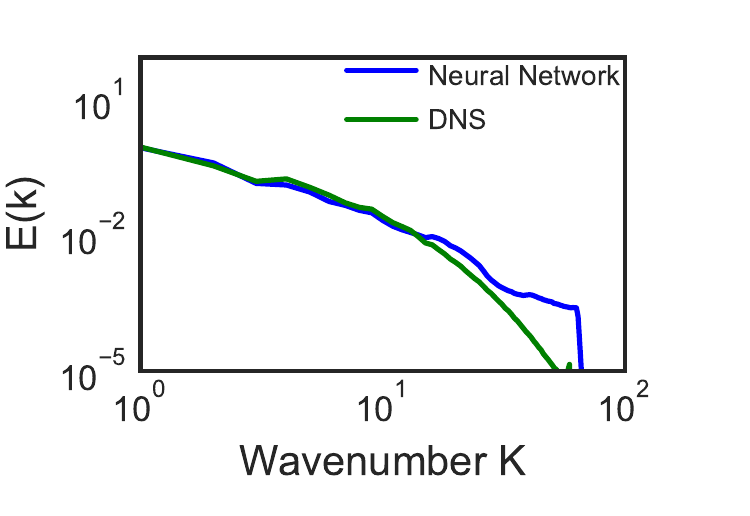}
		\caption{Energy Spectrum}
	\end{subfigure}
	\begin{subfigure}[b]{0.49\textwidth}
	    \centering
		\includegraphics[width = \textwidth]{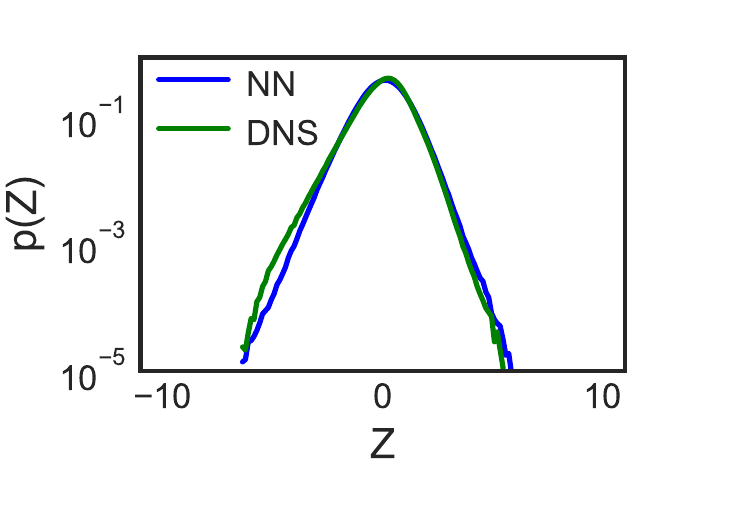}
		\caption{PDF of the longitudinal velocity gradient}
	\end{subfigure}
	\par\bigskip
	\begin{subfigure}[b]{.8\textwidth}
	    \centering
		\includegraphics[width = \textwidth]{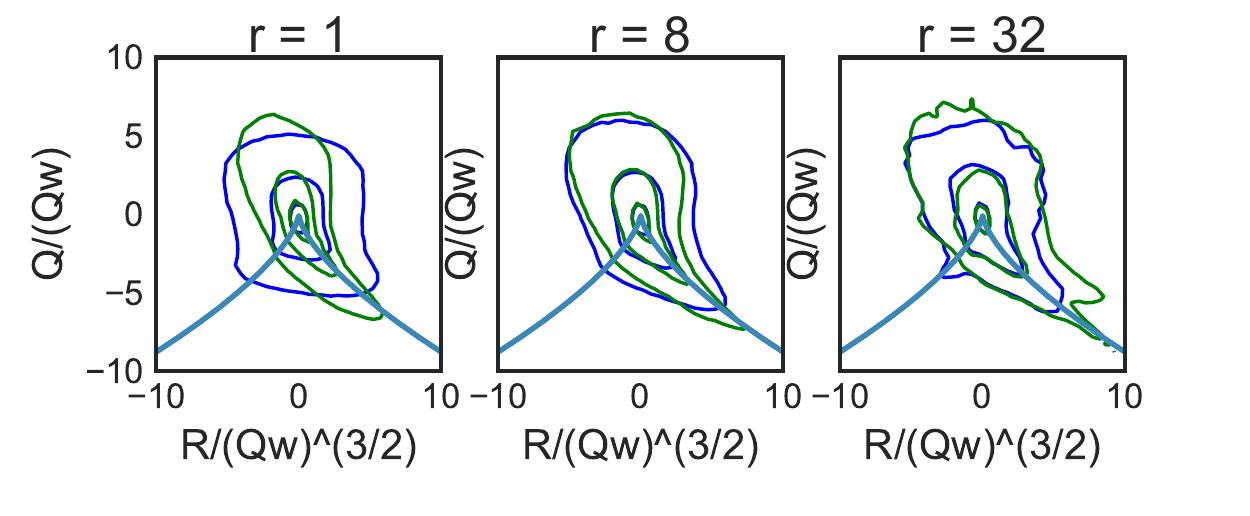}
		\caption{Q-R plane joint PDFs at different scales, where $r$ relates to the grid spacing, such that $r=1$ corresponds to viscous scales, $r=8$ corresponds to inertial range, and $r=32$ corresponds to large scales. The axes are normalized by $Q_w=W_{ij}W_{ij}/2$, where $\mathbf{W}$ is the rotation tensor.}
	\end{subfigure}
	\caption{Physical and statistical diagnostic tests for the Soft Constraint NN implementation.}
	\label{fig_turb_diag_SC}
\end{figure}

Figure \ref{fig_turb_diag_SC} illustrates all three of the turbulence diagnostics mentioned in Section \ref{DCGAN Section}: the energy spectrum, the PDF of longitudinal velocity gradients, and the Q-R joint PDFs at three different scales. Each figure shows the averaged results from three samples generated by the NN. As shown in the figure, the ability of the NN to capture relevant flow statistics is relatively unchanged by the addition of the soft constraint. The energy spectra illustrate the model's ability to replicate the large scales, but the model falters at the smaller scales. The velocity gradient PDF shows that DCGAN succeeds in capturing some of the intermittency effects, as tails are relatively well reproduced. In the final diagnostic, we plot $Q-R$ joint PDFs corresponding to $r$ values that are associated with different scales of the flow. 
The well-known tear-drop shape of the probability isolines becomes more prominent with decrease of the observation scale $r$. As explained in section \ref{DCGAN Section}, we study the $Q-R$ plane based on the velocity gradient filtered at different scales $r$, to examine large ($r=32$), inertial ($r=8$), and small scale ($r=1$) behaviors~\cite{chertkov1999lagrangian}. This allows us to selectively analyze the accuracy of our predictions at different scales, since we are interested in modeling primarily the large and inertial ranges. The $Q-R$ plane PDFs reaffirm that the NN captures between large and inertial scales, although the classic `teardrop' shape seen in literature is marginally less accurate in the inertial scale.   

\subsection{Hard Constraints: Finite Difference Embedding}

\begin{figure}[ht]
	\centering
	\includegraphics[width=100mm]{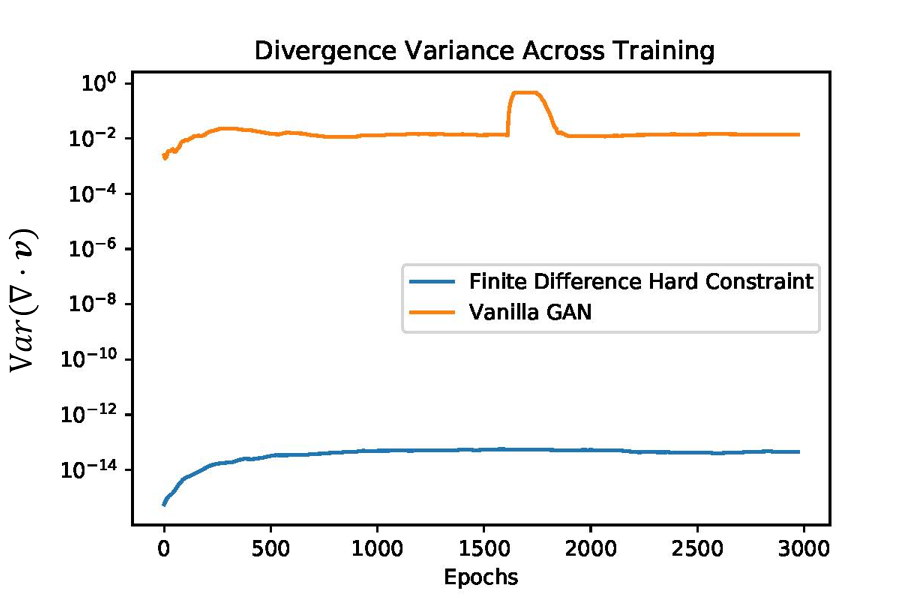}
	\caption{Divergence variance plotted for each training epoch of the finite difference hard constraint embedded GAN.}
	\label{DivFD}
\end{figure}

The first hard constraint method involves embedding a finite difference curl operation directly into the generator network. The results of this approach are seen in Figure \ref{DivFD}, where the hard constraint greatly improves the accuracy to which the incompressibility condition is enforced. In fact, the physics-embedded GAN has reached machine precision with over nine orders of magnitude improvement over the Vanilla GAN. Furthermore, compared to the Vanilla GAN, the embedded GAN shows no instabilities in training.

Notably, the physics embedding hardly increases the training time of the model. With only a 4.7\% increase in run time per epoch, the additional computational cost is negligible and may even be counteracted by quicker convergence of the model. 

\begin{figure}
	\centering
	\begin{subfigure}[b]{0.49\textwidth}
	    \centering
		\includegraphics[width = \textwidth]{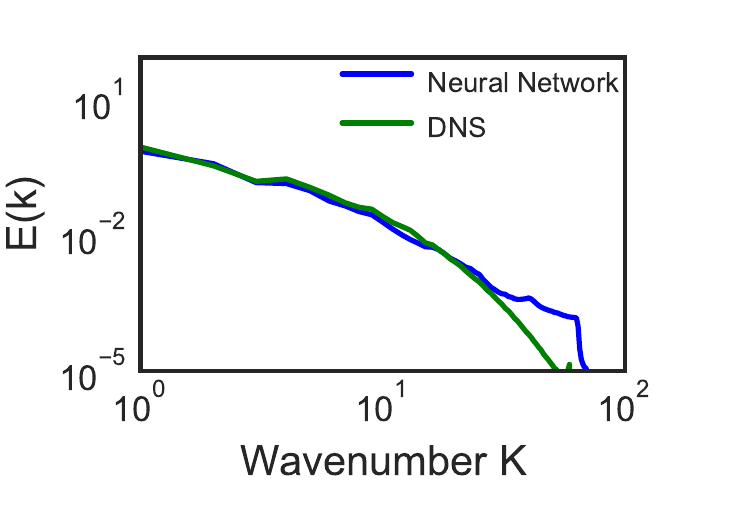}
		\caption{Averaged Energy Spectrum}
	\end{subfigure}
	\begin{subfigure}[b]{0.49\textwidth}
	    \centering
		\includegraphics[width = \textwidth]{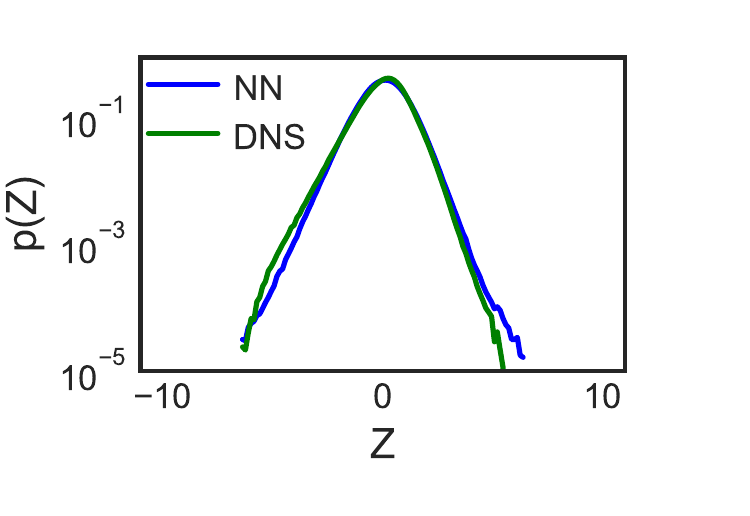}
		\caption{PDF of the longitudinal velocity gradient}
	\end{subfigure}
	\par\bigskip
	\begin{subfigure}[b]{.8\textwidth}
	    \centering
		\includegraphics[width = \textwidth]{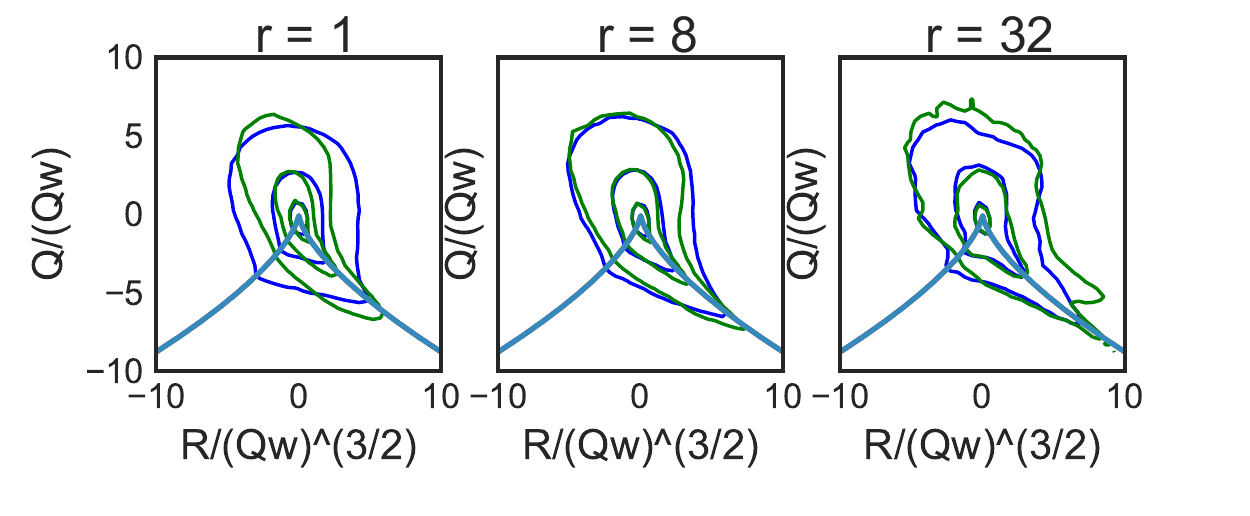}
		\caption{Q-R plane joint PDFs, where $r$ relates to the grid spacing, such that $r=1$ corresponds to viscous scales, $r=8$ corresponds to inertial range, and $r=32$ corresponds to large scales. The axes are normalized by $Q_w=W_{ij}W_{ij}/2$, where $\mathbf{W}$ is the rotation tensor.}
	\end{subfigure}
	\caption{Physical and statistical diagnostic tests for the FD Constraint NN implementation.}
	\label{fig_turb_diag_FD}
\end{figure}

Once again, Figure \ref{fig_turb_diag_FD} shows the resulting three statistical diagnostics from a sample data produced by the FD model. While largely similar to the soft constraints results, there is a noticeable difference in the threshold wavenumber at which the energy spectrum deviates. Furthermore, the large and especially inertial scales are better represented by the model in the Q-R joint PDF plots. 

\subsection{Hard Constraints: Spectral Embedding}

For consistency, this method uses spectral derivatives to calculate the divergence whereas the previous method used finite difference methods. The spectral model only showed an increase of 2.4\% runtime per epoch over the Vanilla Model due to highly optimized, CUDA compliant, Fast Fourier Transform methods in the PyTorch library. 

\begin{figure}[h]
	\centering
	\includegraphics[width=100mm]{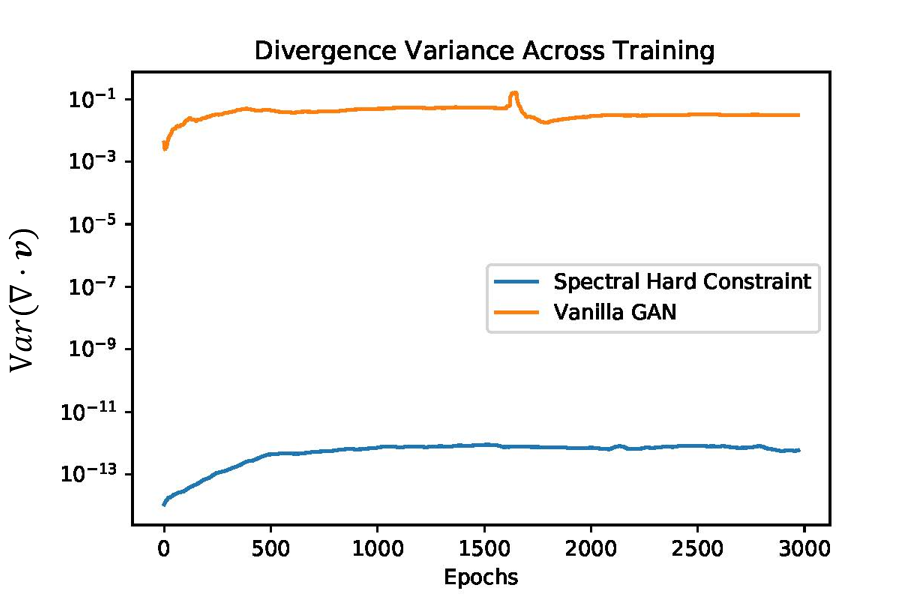}
	\caption{Divergence variance plotted for each training epoch of the spectral hard constraint embedded GAN.}
	\label{DivSpectral}
\end{figure}

We now examine the results for the spectral hard constraint method. Figure \ref{DivSpectral} tracks the divergence of the flow during training. Much like the finite difference method, the spectral method shows an accuracy increase of multiple orders of magnitude over the Vanilla GAN model. The divergence during training is stable and contains no discontinuities.

\begin{figure}
	\centering
	\begin{subfigure}[b]{0.49\textwidth}
	    \centering
		\includegraphics[width = \textwidth]{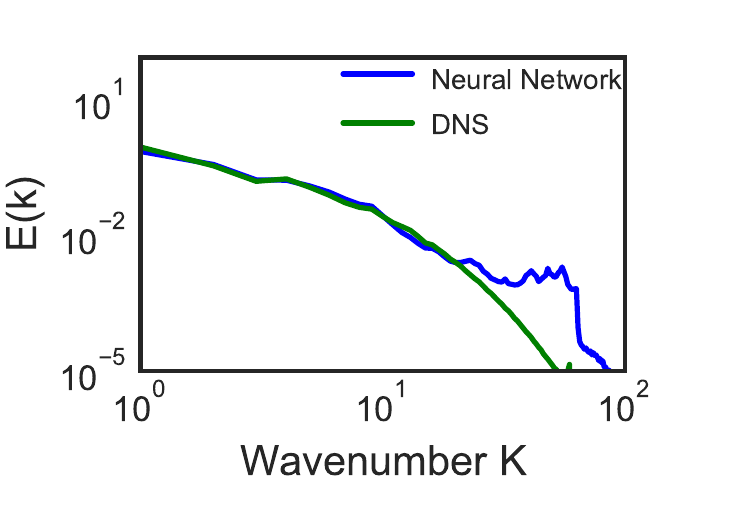}
		\caption{Averaged Energy Spectrum}
	\end{subfigure}
	\begin{subfigure}[b]{0.49\textwidth}
	    \centering
		\includegraphics[width = \textwidth]{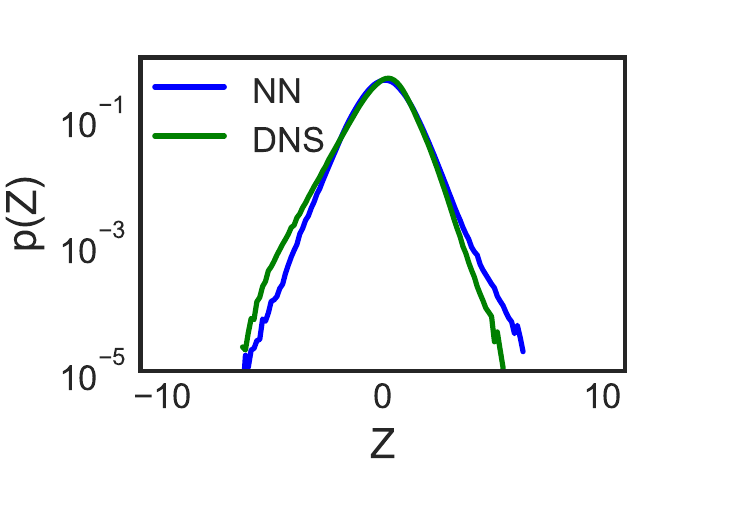}
		\caption{PDF of the longitudinal velocity gradient}
	\end{subfigure}
	\par\bigskip
	\begin{subfigure}[b]{.8\textwidth}
	    \centering
		\includegraphics[width = \textwidth]{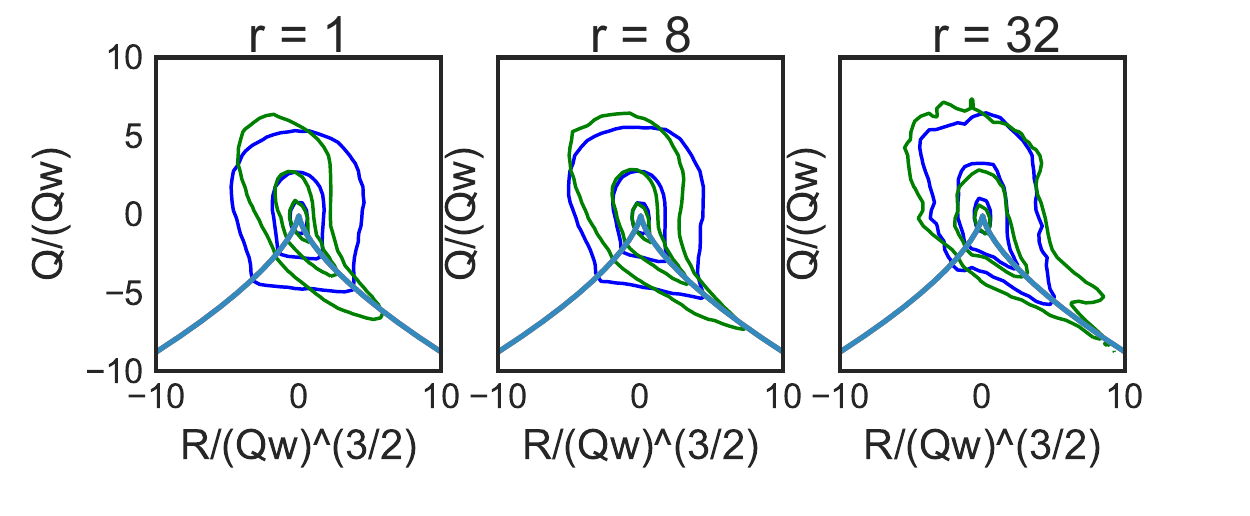}
		\caption{Q-R plane joint PDFs, where $r$ relates to the grid spacing, such that $r=1$ corresponds to viscous scales, $r=8$ corresponds to inertial range, and $r=32$ corresponds to large scales. The axes are normalized by $Q_w=W_{ij}W_{ij}/2$, where $\mathbf{W}$ is the rotation tensor.}
	\end{subfigure}
	\caption{The 3 physical and statistical diagnostic tests for the Spectral Constraint NN implementation.}
	\label{fig_turb_diag_Spec}
\end{figure}

However, the diagnostics results in Figure \ref{fig_turb_diag_Spec} display some key differences compared to the FD method. The energy spectrum diagnostic is comparable to the FD method, with accurate resolution in large and inertial scales while smaller scale features are ignored. However, a discrepancy emerges when we analyze the Q-R plane joint PDFs. Particularly, we notice that the 3D large scale features are not captured well in the spectral method, with considerable deviation from the DNS at larger wavenumbers. Furthermore, the inertial scales, which were previously captured accurately by the FD method, are poorly modeled in the spectral method. This shows that while the spectral method can enforce divergence constraint with similar fidelity as FD, the quality of prediction is comparatively inferior. 

\subsection{Direct Comparison of Hard Constraints Methods}
While there is a substantial improvement in using hard constraints over soft constraints, there is still an important distinction in choosing what type of hard constraints method to use. In this work, we propose two different approaches: based on the finite difference and spectral differentiation. Although Figures \ref{DivFD} and \ref{DivSpectral}, show similar levels of improvement between the two methods, these figures show the variance of the divergence. While an important metric in evaluating the local divergence, it is more revealing in this case to study the mean divergence. For the fully trained networks, the mean divergence of samples from the finite difference version reached $10^{-4}$ while the spectral version reached $10^{-11}$. As a matter of fact, since these are hard constraints methods, they do not need to be trained to enforce the constraint. Hence, at all points in training, the spectral version produces fields for which the mean divergence is multiple orders of magnitude lower than that of the finite difference version. On the surface, this would suggest that the spectral embedding method is superior for adherence to physics constraints. 

However, even though divergence is a constraint of interest, the accuracy of the model is ultimately important. Alongside the continuity Equation, turbulent flow is also governed by the momentum Equations. Although we strictly enforce the divergence constraint, the networks must learn  other flow characteristics from the data since we are unable to impose all constraints dictated by the Navier-Stokes equations. Thus, the turbulence diagnostics shown in Figures \ref{fig_turb_diag_SC}, \ref{fig_turb_diag_FD}, and \ref{fig_turb_diag_Spec} serve as a means to evaluate the ability of the model to learn the overall statistics of the flow. From these diagnostics, it is clear that the spectral constraint method partially impedes the learning of the structure of the velocity field. Hence, there is a clear sacrifice between learning the flow statistics and enforcing the differential constraint. 

\section{Conclusion and Future Work}

In conclusion, we have developed and analyzed two methods for enforcing hard physical constraints in GANs.  While soft constraints are easy to implement and can be effective, they can also lead to noisy results. Hard constraints, on the other hand, are more specialized to implement but lead to more robust and interpretable results. Lastly, we have observed that if the constraint method should be consistent with the goals of the model. For instance, if the model must impose mass conservation over everything else, then the spectral method would be a wise choice. On the other hand, if the other flow statistics are also important, then the FD method outperforms. 

In future work, this embedded GANs platform can be expanded via recurrent LSTM or GRU layers to enable full spatio-temporal forecasting capabilities. The physics embeddings are compatible with a variety of different architectures, due to the ubiquity of convolutional networks in deep learning. Furthermore, the constraint methods can be further improved upon to enforced more generalized differential constraints present in the Navier-Stokes Equations. This method imposes constraints of form $L(V) = G$, where $L$ is any linear differential operator. Therefore its not restricted to fluid dynamics and be extended to problems in other fields where PDEs are prevalent. As our understanding of DL continues to progress, physics-informed methods have the potential to become very effective at low evaluation costs. 

Generative modeling, specifically, has evolved drastically in recent years, owing to highly realistic 2D models such as \cite{Brock} and \cite{Karras}. Taking a similar approach to \cite{Mohan_JoT}, these 2D methods can be extended to 3D in the hopes of resolving smaller turbulent structures. But GANs are just the tip of the iceberg; Transformers \cite{Vaswani}, for instance, have quickly become immensely popular and effective in modeling time series data. Combining these architectures with time-tested numerical and spectral methods opens the door to powerful models not achievable solely through DL. 

Overall, the paradigm of physics-informed ML leverages decades of accumulated knowledge of physical phenomena in order to improve data-based models for better convergence, extrapolation, and accuracy. While we have shown the considerable effectiveness of hard constraints, there are still improvements to be made in terms of generalization to a wider array of operators. In particular, our work shows that the rich history of numerical methods can also aid in deep learning for PDEs, without incurring additional penalty in computational costs.

\printbibliography

\newpage
\begin{appendix}
\section{Vanilla GANs Structure and Training} \label{GANs Details}
In the following, we discuss the details of the Vanilla GAN, the base GAN for all three constraint implementations. Both the Discriminator and Generator Networks were trained with a feature map size of $64$ and with a batch size of $12$. The noise vector to initialize the training was of size $100 \times 1$. And the ADAM optimizer \cite{Kingma} with optimization parameters $\beta_{1} \,=\, 0.5$ and $\beta_{2} \,=\, 0.999$ was used to train both network. The learning rates are discussed in Section \ref{lr}. 

\subsection{Transpose Convolution and Resize Convolution}\label{RC}

\begin{figure}[ht]
\includegraphics[width=10cm]{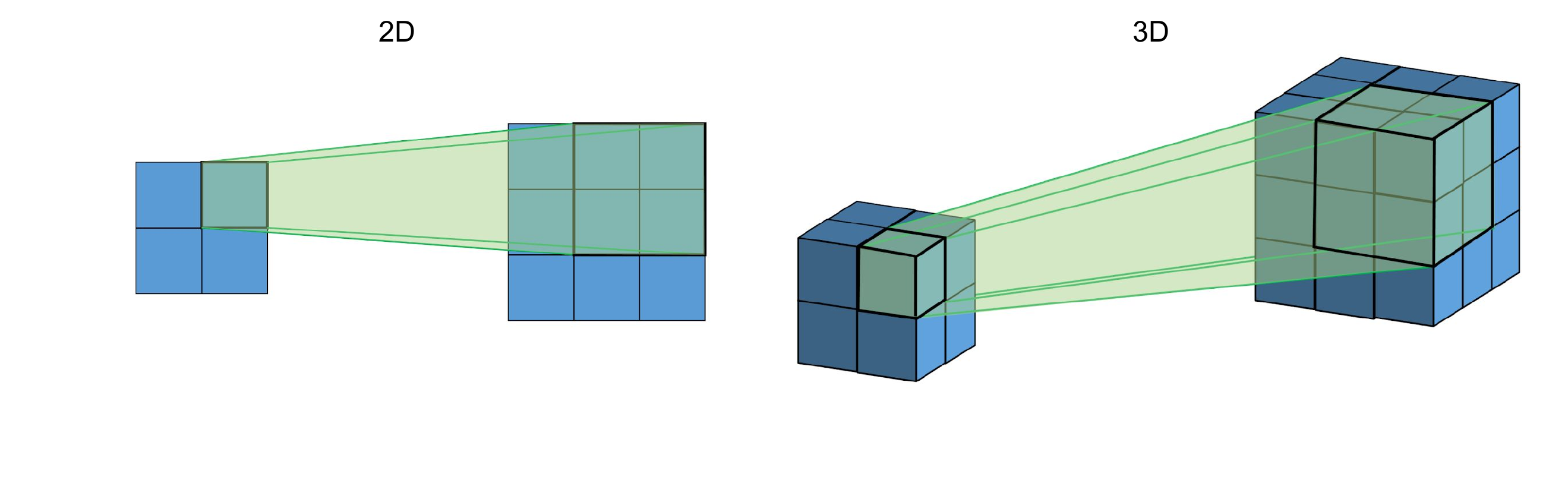}
\centering
\caption{Representation of transpose convolution. In this case, an input size of 2 is up sampled to an output size of 3 with a kernel size of 2 (square kernel for 2D and cubic for 3D) and a stride of 1. 2D transpose Convolution is shown on the left and volumetric transpose convolution is shown on the right.}
\label{3DConv}
\end{figure}

Transpose convolutions are the traditional approach to upsampling used in CNN based GANs in the Generator Network. This operation can be thought of as the reverse of a standard convolution: instead of sliding a kernel across a group of pixels to learn a mapping to fewer pixels, the kernel is trained to extrapolate individual pixels to a larger pixel group. The distance that the kernel slides each time is known as the stride. Deconvolution is sometimes mentioned as the same operation even though the two operations are not the same (deconvolution is, technically, the inverse of convolution). Since we are dealing with volumetric data, we utilize 3D transpose convolutions which use a cubic kernel. Furthermore, the padding determines how many layers of zero-value pixels are added to the edges of the input. See Figure \ref{3DConv} for a comparison of 3D and 2D transpose convolution. The use of transpose convolutions in the Generator results in a common issue known as checkerboard artifacting~\cite{Odena}. The artifacts are the result of overlapping transpose convolutions when upsampling the data. This typically occurs when the stride is less than the kernel size, especially when the kernel size is not divisible by the stride.

Odena et al.\cite{Odena} provide an interesting solution to the checkerboard artifact problem, known as resize convolution (RC). RC involves interpolation followed by standard convolution. We found that trilinear interpolation worked best for our application whereas nearest-neighbor interpolation continued to result in some line artifacts. Trilinear interpolation consists of inserting zero padding in between values in the input tensor (to resize the sample to the desired dimensions) followed by averaging the values close to the padding to determine the new value of those indices. We also found that the Generator does not learn when solely using RC. To determine if the Generator network was learning, we suspend updates to the Discriminator, and continue training the Generator. With the discriminator no longer learning, the generator would have no competition and therefore should begin to learn to output samples that the discriminator will classify as “real”. However, if the discriminator still continued to identify the generated images as “fake”, then we can concur that the generator was not learning. This was indeed the case with the RC-only Generator above.

\begin{figure}[ht]
\includegraphics[width=10cm]{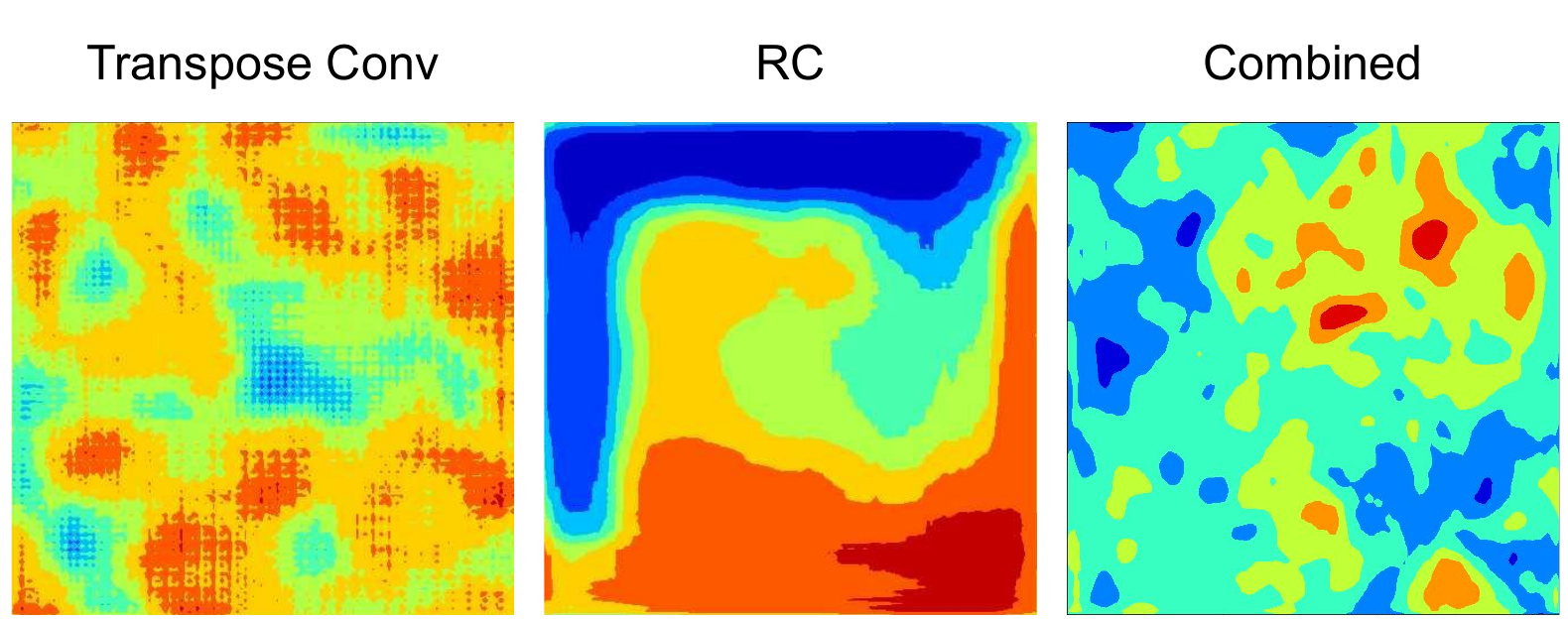}
\centering
\caption{An illustration of using only transpose convolution (left), only resize convolution (middle), and the transpose-RC hybrid (right) on a 2D slice of the flow}
\label{TC_RC}
\end{figure}

Instead of choosing between the two approaches, we employ a hybrid strategy with transpose convolution in the first few layers of the Generator and RC in the rest of the layers. This scheme proved successful as the transpose convolutional layers learned the underlying distribution of the data, while the RC layers learned to smooth out and eliminate the checkerboard artifacts. Figure \ref{TC_RC} illustrates the results of using only one method of upsampling followed by combining both methods.

\subsection{Label Smoothing} \label{label_smoothing}
In GANs, the Discriminator has a tendency to become overconfident and outputs labels very close to 1 or 0 (as opposed to labels such as 0.9 or 0.1). When this happens, the Generator’s gradients begin to vanish and it can no longer meaningfully update its weights. One solution to this is to attempt to balance the networks by making the Generator "stronger" or the discriminator "weaker". Although this solution can be effective, we found it ultimately limited the potential of the GAN. Therefore, we implement a solution first mentioned by Salimans\cite{Salimans}: one sided label smoothing. Instead of using 1 and 0 as target labels, we added noise to the labels so that the real label would fall in the range $[0.875, 1]$, and the fake label would fall in the range $[0, 0.125]$. This effectively decreased the Discriminator’s overconfidence so that it could still accurately classify the samples without compromising on the gradients it provides to the Generator.

\subsection{Cyclic Learning Rates}\label{lr}
\begin{figure}[ht]
\includegraphics[width=10cm]{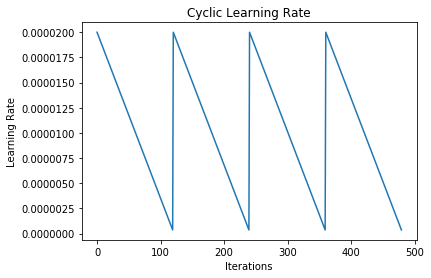}
\centering
\caption{Cyclic Learning Rate: Triangular Update Policy}
\label{cyclic lr}
\end{figure}

In order to improve training efficiency, we employed the technique of cyclic learning rates by Smith \cite{Smith} . Varying the learning rate as the models train, allows them to converge faster and reach a lower loss value. Figure \ref{cyclic lr} shows how our learning rate varied as the models trained. We employed a triangular update policy, changing the learning rate in a piece-wise linear fashion. Hence, the learning rate fluctuates between minimum and maximum values at a rate determined by the number of steps it takes to complete one full cycle. In this work, the min and max rates were set as $2 \times 10^{-7}$ and $2 \times 10^{-5}$ respectively. Although these learning rates are rather small, we found that learning rates set any higher would destabilize the network. Changing the learning rate every iteration allowed us to forsake finding a perfect value for a constant learning rate. Although Ref.~\cite{Smith} detailed an excellent way to find the minimum and maximum values for a given classification model, GANs converge in a different manner that is not entirely clear from losses alone. Therefore, to select these values we found the minimum and maximum values at which the losses did not diverge, but also learned at an acceptable speed. Furthermore, the Discriminator's learning rate is an order of magnitude less than the Generator, in order to balance the networks' relative strength. 

\end{appendix}
\end{document}